\documentclass{pinchcr}   

\usepackage{graphicx}        
\usepackage{amsmath}

\renewcommand{\phi}{\varphi}

\newcommand{\be}{\begin{equation}}
\newcommand{\ee}{\end{equation}}

\title{The Length Scales of Dynamic Heterogeneity: Results from Molecular Dynamics Simulations}

\author{Peter Harrowell}
\affiliation{School of Chemistry, University of Sydney, Sydney NSW 2006, Australia.}

\begin{document}

\maketitle

\preface
Over times shorter than that required for relaxation of enthalpy, a liquid can exhibit striking heterogeneities. The picture of these heterogeneities  is complex with transient patches of rigidity, irregular yet persistent, intersected by tendrils of mobile particles, flickering intermittently into new spatial patterns of motion and arrest. The study of these dynamic heterogeneities has, over the last 20 years, allowed us to characterize cooperative dynamics, to identify new strategies in controlling kinetics in glass-forming liquids and to begin to systematically explore the relationship between dynamics and structure that underpins the behaviour of amorphous materials. Computer simulations of the dynamics in atomic and molecular liquids have played a dominant role in all of this progress. While some may be uneasy about this reliance on modelling, it is unavoidable, given the amount of microscopic detail needed to characterize the dynamic heterogeneities. The complexities revealed by these simulations have called for new conceptual tools. In this essay, I have tried to provide the reader with a clear and complete account of how these tools have been developed in terms of the literature on kinetic length scales in molecular dynamics simulations. Through the `prism' of these length scales, this essay addresses the question what have we learnt about dynamic heterogeneities from computer simulations?

\section{Introduction}
\label{intro}

What is it that distinguishes a glass from a crystal? Starting with the most casual inspection, the presence of oriented planes, grain boundaries or edges
would indicate a crystal since the glass must be isotropic (and, hence, amorphous).  Looking closer, the presence of sharp intensity peaks in scattered
radiation at large angles indicates the presence of repeated parallel planes of density. The absence of such planes is typically all it takes for us to
label the material `disordered'. At the level of the constituent particles (atoms or molecules), however, the most striking physical consequence of being
in a glass rather than a crystal is the large number of different local environments (i.e. structural heterogeneity) in the former. Daama and
Villars~\shortcite{daama97} have established that over 90\% of the known 17000 inorganic crystal structures with intermetallic structure types consist of
no more than 4 distinct coordination environments. This is in clear contrast to a corresponding glass. Reverse Monte Carlo analysis of EXAFS measurements of the intermetallic glass $\mathrm{Ni}_{80}\mathrm{P}_{20}$~\shortcite{luo08} have identified over 15 different coordination environments. Collective dynamics serves to amplify these structural variations, translating the often subtle differences in local configurations
into dramatic variations in local relaxation kinetics. The result is that the approach to the glass transition is typically marked by striking spatial
heterogeneity in dynamics. Many of the landmark insights of material science - dislocation-mediated plasticity, the Peierls barrier, Nabarro-Herring creep,
grain boundary mobility - can be comfortably re-expressed as manifestations of dynamic heterogeneities. In crystals, these objects are identifiable as defects and
therefore have explicit structural signatures but, by shifting our focus to the spatial distribution of dynamics rather than order, we can expand our
study of localised relaxation processes to materials for which we have no a priori notion of the structural origin of the localisation. From this
perspective, dynamic heterogeneities provide a genuine opportunity to develop a universal description of dynamic localization and collective relaxation in
condensed matter - ordered and disordered, in equilibrium or out.

Over the last 20 years, there has been a steady growth in the appreciation of the ubiquity of dynamic heterogeneities in disordered materials and of the
value of their study. The subject of dynamic heterogeneities has been considered in a number of reviews of the glass transition
~\shortcite{poole98a,kob99,glotzer00a,jackle02,andersen05,binder05,heuer08}. These spatial fluctuations, which take the form of transient kinetic domains,
represent an extension of the `traditional' phenomenology of disordered materials (`traditional' referring to thermodynamics and bulk averaged scattering
and dynamic susceptibilities); an extension capable of providing spatial information about the fluctuations associated with the collective dynamics without
requiring any insight as to the particle arrangements responsible. The existence of heterogeneities can also provide a link between different aspects of
the phenomenology - stretched relaxations, fragility, etc.~\shortcite{perera96}. While computational methods may dominate their study and theoretical goals
provide much of the motivation, it is worth emphasising that dynamic heterogeneities are not a theoretical construction but a physical fact and their
description is, in the end, of value in its own right.

To talk about dynamic heterogeneities in anything other than pictures we need some quantities to measure and kinetic lengths represent the most versatile
of these, making them something of a \emph{lingua franca} of glassy dynamics. What do we mean by a kinetic length? Prior to the study of dynamic
heterogeneities, the kinetic length was attributed, somewhat vaguely, to a size of collective motion or a cooperative rearrangement. In this essay we shall
review what we now know about kinetic length scales associated with dynamic heterogeneities as a result of molecular dynamics (MD) simulations. The
restriction to MD simulations, while omitting many important aspects of the research into dynamic heterogeneities, ensures that we shall only consider
those aspects of cooperative dynamics that we can explicitly connect to the positions and momenta of particles. We shall not cover the work on lattice
models of glass forming liquids and we shall not cover the various theoretical treatments of the glass transition. Interested readers can find excellent reviews of these important topics in
refs.~\shortcite{ritort03} and \shortcite{cavagna09}, respectively, and in the chapters of this volume.

We shall take the view that MD simulations confront us with all of the essential complexities of real supercooled liquids and glasses without, necessarily,
providing a quantitatively accurate model of any one specific glass former. The great boon of these simulations, i.e. the provision of all particle
positions and momenta at as many instances as desired, is also the source of the rather stringent obligation that they exert. After all, if you can access
all information, then the failure to explain any aspect of the glass transition that can be captured by a simulation can only be attributed to our personal
failure in asking the right question. Dynamic heterogeneity owes much of its existence as a recognised phenomenon to MD simulation studies. While we now
have mesoscopic~\shortcite{weeks00,abate07} and macroscopic~\shortcite{dauchot05} analogues of liquids in which the dynamic
heterogeneities can be directly seen, it is fair to say that these discoveries owe their validation as models of microscopic heterogeneities to comparisons
with computer modelling.

The idea of a kinetic length scale, which extends back, at least, to the work of Adam and Gibbs~\shortcite{adam65} in 1965 (if not further back~\shortcite{jenckel39}), is that the
growing degree of dynamic correlation in a supercooled liquid can be measured in terms of a length that grows with cooling. To be accurate, Adam-Gibbs did
not talk of a length scale but, rather, a number of particles required for a collective rearrangement. Converting this number into a length requires some
sort of ancillary assumption concerning the connectivity of these particles. There is an undeniable wishfulness involved in expecting to be able to replace
something so poorly defined as the `extent of cooperative motion' with something as concrete as a length. Computer simulations have provided the
opportunity (obligation, really) to put this aspiration to the test and develop explicit expressions (theoretical or algorithmic) for this kinetic length
scale.

MD simulations require interparticle potentials. Simulations of supercooled liquids require interactions that are computationally simple in order to get as
many time steps out of a computer as possible and yet geometrically complex enough to stave off crystallization long enough to allow characterization of
the metastable liquid. In Table \ref{tab:potentials} we provide details of a number of models, all based on binary atomic alloys, that have proved popular
and which form the basis of many of the simulation studies of the glass transition.

In this review we shall consider five approaches to calculating a kinetic correlation length from MD simulations: direct measures of the spatial
distribution of mobility, 4-point correlation functions and the associated susceptibility $\chi_{4}$, finite size effects, kinetic correlations at
interfaces and the crossover lengths for kinetic properties.

\begin{table}
\label{potentials}

\begin{tabular}{cccc}
\bf{Label} & \bf{Inter-particle Potential} & \bf{Composition} & \bf{Reference}\\
\hline
\\
BMLJ1 & $\phi_{ij}(r)=4\epsilon_{ij}([\sigma_{ij}/r]^{12}-[\sigma_{ij}/r]^{6})$ & A$_{4}$B & a \\
\emph{(Kob-Andersen)} & $\sigma_{AA}=1.0$,$\sigma_{BB}=0.88$, $\sigma_{AB}=0.8$ & &  \\
& $\epsilon_{AA}=1.0$,$\epsilon_{BB}=0.5$,$\epsilon_{AB}=1.5$ & & \\
\\
\hline
\\
BMLJ2 & $\phi_{ij}(r)=4\epsilon_{ij}([\sigma_{ij}/r]^{12}-[\sigma_{ij}/r]^{6})$ & AB & b\\
\emph{(Wahnstr\"{o}m)} &  $\sigma_{AA}=1.2$,$\sigma_{BB}=1.0$, $\sigma_{AB}=1.1$ & & \\
&  $\epsilon_{AA}=\epsilon_{BB}= \epsilon_{AB}=1.0$ & & \\
\\
\hline
\\
SS & $\phi_{ij}(r)=\epsilon_{ij}(\sigma_{ij}/r)^{12}$ & AB &  c\\
\emph{(soft sphere)} & $\sigma_{AA}=1.2$,$\sigma_{BB}=1.0$, $\sigma_{AB}=1.1$ & &  \\
& $\epsilon_{AA}=\epsilon_{BB}= \epsilon_{AB}=1.0$ & & \\
\\
\hline
\\
SD & $\phi_{ij}(r)=\epsilon_{ij}(\sigma_{ij}/r)^{12}$ (in 2D) & AB & d\\
\emph{(soft disk)} & $\sigma_{AA}=1.4$,$\sigma_{BB}=1.0$, $\sigma_{AB}=1.2$ & &  \\
& $\epsilon_{AA}=\epsilon_{BB}= \epsilon_{AB}=1.0$ & & \\
\\
\hline
\\
\end{tabular}
References: a)~\shortcite{kob95}, b)\shortcite{wahnstrom91}, c) \shortcite{bernu87}, d) \shortcite{perera98a}\\
\\
\caption{\label{tab:potentials} Details of a number of model glass forming liquids. Note that BMLJ1 and BMLJ2 differ in that the former includes a strong
preference for AB neighbours while the latter does not impose any specific chemical ordering. The soft sphere (SS) model is the same as the BMLJ2 system
except for the absence of the attractions.}

\end{table}

\section{Kinetic Lengths From Displacement Distributions}
\label{displacements}

Historically, the first approach~\shortcite{deng89} to analysing dynamic heterogeneities was to make a map of them. Maps are appealing. They retain a large
amount of information. A single kinetic length scale, as we shall see, represents a major (and often uncontrolled) discarding of most of this information.
The appeal of a map must be weighed against their principal short coming - they provide information (a lot of it, admittedly) about a single instance of
heterogeneity only. Considerable care, therefore, is generally required in extracting the statistically significant features from the noise.

\begin{figure}[t]
\centering
\includegraphics*[width=8 cm]{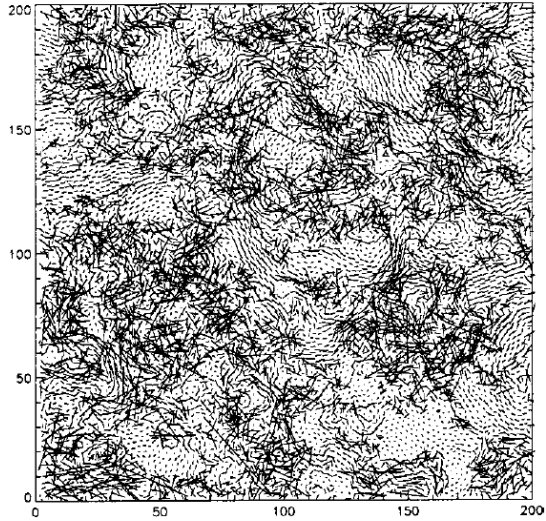}
\caption[]{The particle motions $\vec{r}_{i}(t+50\tau_{\alpha})-\vec{r}_{i}(t)$ where $\tau_{\alpha}$ is the structural relaxation time for a polydisperse
mixture of hard disks. [Reproduced with permission from ref.~\shortcite{doliwa00}. ]
}
\label{figone}       
\end{figure}

The earliest example of displacement maps being used to specifically characterise dynamic heterogeneities can be found in a 1989 paper by Deng, Argon and
Yip~\shortcite{deng89}. This paper contains a section with the prophetic title \emph{Inhomogeneities and the clustering of atomic motions}. Inspection of
maps of the particle displacements over a chosen time interval (see Fig.~\ref{figone} for an example), reveal the generic features of the heterogeneities. Typically,
we find connected domains of slow and of fast particles, with the former typically more compact than the latter. Collective strains within the `slow'
domains coexist with low dimensional flows among the more mobile particles.  Forced to identify lengths we might choose the average extent of slow domains
or the fast domains or we might consider the length scale over which displacement direction is correlated. Alternatively, we might ask about the size of
the `core' regions where large displacements appear to be directed randomly or, instead, determine the average separation between such localised
reorganizations. Our choice would, of course, be simplified if an argument existed that established that all of these lengths scaled in a similar fashion.
Unfortunately, no such argument exists.

In 1995, Hurley and Harrowell~\shortcite{hurley95} extracted a kinetic length scale for the one component soft disk liquid in terms of the decay in the
variance of relaxation time maps as the linear dimension of the scaling volume increased using a box scaling method. (That these calculations were carried
out on an equilibrium liquid - and only later on a supercooled mixture~\shortcite{perera98a} - underscores the point that dynamic heterogeneities are not
restricted to supercooled liquids.) The relaxation time was defined as the first passage time for a particle's displacement to exceed a threshold distance
that was chosen to maximise the kinetic length. The kinetic length in the SD liquid was found to exhibit a super-Arrhenius increase with decreasing
temperature~\shortcite{perera98a,perera98b,perera99}.

In 1997 Kob et al~\shortcite{kob97} introduced an analysis of dynamic heterogeneities based on determining the statistics of clusters of various kinetic
subpopulations. Mobile particles were defined in such a way that they comprised of the  $\sim$ 5\% of particles with the largest displacements over a time
interval corresponding to the maximum in the non-Gaussian parameter (see Section \ref{crossover}). A characteristic size was obtained as S, the mass
weighted average cluster size. For the BMLJ1 liquid, the temperature variation of S was fitted as $S = 0.975/(T-0.431)^{0.687}$~\shortcite{donati99a}.
Previously, the mode coupling theory~\shortcite{gotze92} had predicted the divergence of the relaxation time via a similar functional form, i.e.
$(T-T_{c})^{-\zeta}$. In the BMLJ1 liquid, $T_{c} \sim 0.43$~\shortcite{kob97}. The analogous average cluster size for the 5\% \emph{slowest} particles
exhibited little variation with T. The authors suggested that the apparent divergence of the size of the mobile clusters was significant, possibly linked
to the divergence predicted by the mode coupling theory [32]. While an increase in the kinetic length with cooling is confirmed by all approaches, some
care needs to be taken in interpreting an increase in cluster size when one is looking at clusters of some subpopulation of fixed size. A decrease in the
total number of such clusters will force an increase in the size of those remaining, simply as an artefact of how the clusters are defined.

The cluster analysis has been applied to a number of systems: polymers~\shortcite{bennemann99,gebremichael01}, water~\shortcite{mazza06},
SiO$_{2}$~\shortcite{vogel04a,vogel04b}, the Dzugutov potential~\shortcite{gebremichael04} and polar diatomic molecules~\shortcite{palomar08}.
Giovambattista et al~\shortcite{giovambattista05} have calculated the fractal dimension of the mobile clusters in supercooled water. They found that large
clusters exhibited a fractal dimension of $\sim$ 2, a value similar to that predicted~\shortcite{lamarcq02} for low energy excitations in a spin glass.
Vollmayr-Lee et al~\shortcite{vollmayr-lee02} have extended the mobile cluster analysis to a BMLJ1 mixture \emph{below} its glass transition temperature.
Below T$_{g}$, mobility is determined using the mean amplitude of a particle's fluctuation about its mean position. The authors report that clusters of
mobile particles defined by this means were more compact and shorter lived than the analogous clusters above T$_{g}$. As to the question of the connection
between the kinetic lengths of mobile particles on either side of the glass transition, strong correlations have been demonstrated~\shortcite{widmer-cooper06} above T$_{g}$ in
the SD mixture between the spatial distribution of mobile particles and those particles which, over short times, exhibited large amplitude fluctuations
(analysed in terms of local Debye-Waller factors~\shortcite{widmer-cooper06}) - a criterion similar in spirit to that used by Vollmayr-Lee et
al~\shortcite{vollmayr-lee02}.

Having introduced the cluster analysis, Donati et al~\shortcite{donati98} augmented it in order to examine the correlation between displacement directions
and the positions of mobile neighbours. Starting with the same fraction of mobile particles, they introduced an additional requirement for belonging to a
cluster - for two neighbouring mobile particles to belong to the same cluster the new position of one particle has to lie within a selected distance of the
neighbours old position. This new overlap condition strengthens the interpretation of the clusters as representing a cooperative motion rather than simply
some general aggregation. Striking pictures of 'string-like' arrangements of displacements (i.e. collective movements that are correlated only along the
direction of flow) have been presented for the BMLJ1~\shortcite{donati98} and SS~\shortcite{kim00} mixtures. The observation of low dimensional particle
flows raises some interesting questions about the mechanisms of cooperative dynamics and readers are encouraged to read
refs.~\shortcite{vogel04a,gebremichael04} closely for details of the complex motions by which displacement strings are formed.

Since first presented in 1998~\shortcite{donati98}, images of strings of particle displacements have proved a popular \emph{leif-motif} for dynamic
heterogeneities in general. To what degree is this representation accurate? Are dynamic heterogeneities generally string-like or are extended strings of
displacements simply the more eye catching members of a broad distribution of cluster shapes? There is not a lot of information that addresses this
specific question. Focusing on the most mobile $\sim$ 5\% of particles in a BMLJ1
mixture, Donati et al~\shortcite{donati98} demonstrated that the displacement vectors exhibited a local forward alignment and that the clusters defined
using the overlap condition were described by an exponential distribution of the size with the
average number increasing from $\sim$ 1.4 (T = 0.55) to $\sim$ 2.2 (T = 0.45). The mass weighted average size is larger, reaching $\sim$ 15 at T =
0.45~\shortcite{donati99a}. While the authors of ref.~\shortcite{donati99a,donati98} refer to these clusters as `strings', the overlap condition for
cluster membership establishes only a flow and does not specifically establish the dimensionality of that flow. While the overlap condition will certainly
include string-like correlations, it will also include some collective flows and strains in general. The existence of such collective flows in liquids at
rest has been known for some time. Alder and Wainwright~\shortcite{alder67} demonstrated in 1967 the local forward alignment between particle velocities in
an equilibrium liquid, part of a pattern that strongly resembled the solution of the Navier-Stokes equations about a moving particle. Doliwa and
Heuer~\shortcite{doliwa00} have reproduced this pattern in a study of the displacements in a binary hard disk mixture. To establish string-like
correlations one would like to establish the average coordination number of particles in the mobile cluster is $\le 2$. This quantity has not been
determined but Donati et al~\shortcite{donati99a} have reported a fractal dimension of 1.75 for mobile clusters in the BMLJ1 model (obtained over a single
order of magnitude data set).  There are indisputably string-like objects such as the self avoiding random walk in 3D with a similar fractal
dimension (1.66 from Flory's argument~\shortcite{flory69}). There are, however, other random objects with similar fractal dimensions - on a 3D lattice we have
the random cluster below percolation with a fractal dimension of 2 and the backbone of the percolating cluster with a dimension $\sim$ 1.7~\shortcite{stauffer94}
- for which the description `string-like' really doesn't apply.  The cumulative data clearly points to mobile clusters that are \emph{not} compact 3D
objects but it leaves open the question of what they \emph{are}. There is certainly a need for more systematic studies of the shape (and its dependence on
temperature and the time interval used to define mobility) as well as the length scale of dynamic heterogeneities.

In simulations of the SD and SS mixtures, Yamamoto and Onuki~\shortcite{yamamoto97,yamamoto98a,yamamoto98b,yamamoto99} shifted the focus from particle
displacements to changes in the topology of a configuration in the form of `broken bonds'. Their analysis involved generating spatial maps of the positions
where initially nearest neighbour pairs first moved far enough apart for the two particle `bond' to be considered broken. These authors showed that the
structure factor S$_{b}$(q) of the broken bonds in both 2D and 3D, accumulated over a fixed time interval, obeyed the small q expansion of $S_{b}^{-1}(q)$,
i.e.

\begin{equation}
\label{one}
S_{b}(q)\approx \frac{S_{b}(0)}{1+\xi^{2}q^{2}}\,
\end{equation}

\noindent where $\xi$ represents the associated length. (The functional form in Eq.~\ref{one} corresponds to an exponential decaying correlation between broken
bonds in real space.) Observing that S$_{b}$(q) appeared independent of temperature (and, hence, $\xi$) at large q, the authors concluded that $S_{b}(0) \sim
\xi^{2}$, the Ornstein-Zernicke result, and noted the analogy with critical fluctuations.

Analogous correlation functions have also been developed to describe the spatial correlations between displacements. In 1998, Poole et
al~\shortcite{poole98b} introduced the displacement-displacement correlation function $G_{u}(\vec{r},\Delta t)$, defined as follows. For the displacement amplitude field
$u(\vec{r},t,\Delta t)$ given by

\begin{equation}
u(\vec{r},t,\Delta t)=\sum_{i=1}^{N}\vert\vec{r}_{i}(t+\Delta t)-\vec{r}_{i}(t)\vert \delta(\vec{r}-\vec{r}_{i}(t))
\end{equation}

\noindent we can write

\begin{equation}
\label{two}
G_{u}(\vec{r},\Delta t)=\int d\vec{\acute{r}}<[u(\vec{\acute{r}}+\vec{r},t,\Delta t)-<u>][u(\vec{\acute{r}},t,\Delta t)-<u>]>
\end{equation}

By analogy with the relationship between the density-density correlation function, the variance of the number of particle N and the compressibility $\kappa$, Donati et al~\shortcite{donati99b} write

\begin{equation}
\label{three}
\int d\vec{r}G_{u}(\vec{r},\Delta t)=<[U-<U>]^{2}> \equiv <u><U>kT\kappa_{u}
\end{equation}

\noindent where $U=\int d\vec{r}u(\vec{r},t,\Delta t)$ is the total displacement and $\kappa_{u}$ is a time dependent dynamic susceptibility. For the BMLJ1 model, the maximum value of $\kappa_{u}$ with respect to the displacement time interval $\Delta t$ varies with T as $(T-0.435)^{-0.84}$. From the decay of the density correlations with separations or, equivalently, working with the analogous structure factor and Eq.~\ref{one} as in ref.~\shortcite{yamamoto98a}, a kinetic length can be extracted. This kinetic length corresponds to the correlation length of fluctuations in displacement amplitude field. A number of models have been analysed using this approach: the BMLJ1 mixture~\shortcite{poole98b,donati99b}, molecular~\shortcite{qian99} and polymeric~\shortcite{bennemann99,gebremichael01} liquids and polydisperse hard spheres~\shortcite{doliwa00}.

The connection between length scales and relaxation times represents the main motivation for looking at the length scales in the first place. Empirical power law relations between a relaxation time and the length of the form

\begin{equation}
\label{four}
\tau = A \xi^{z}
\end{equation}

\noindent were reported in all of the above cases, often with large exponents. In the two dimensional SD mixture $z \sim 4$ from both the box scaling~\shortcite{perera98a} and broken bond structure factor~\shortcite{yamamoto98a}. An even larger exponent is found in the BMLJ1 mixture~\shortcite{donati99a},  where $\tau_{\alpha}\sim S^{4.5}$. Since S is the average (mobile) cluster size, it will be related to a length $\xi$ via $S \sim \xi^{\nu}$, where $1 < \nu < 3$, implying, when substituted back into the power law relation with $\tau_{\alpha}$, a value of $z > 4.5$. The BMLJ2 and SS mixtures, in contrast, have modest exponents: $z \sim 2$ (bond breaking)~\shortcite{yamamoto98a} for the SS mixture and $z \sim 2.34$ (4-point correlation)~\shortcite{lacevic03} in the BMLJ2 model. We remind the reader that the BMLJ2 and SS models have the same short range repulsions and associated length scales. The smaller the value of $z$, the larger the kinetic length required to achieve a given relaxation time. Does this mean that the structures responsible for $\xi$ in these two liquids are mechanically less robust than in the SD and BMLJ1 liquids, since they must be larger to achieve an equivalent stability?

The difference between the value of $z$ for the SS and SD liquids is also noteworthy since both only make use of short range repulsions. They differ, of course, in their spatial dimension. Yamamoto and Onuki~\shortcite{yamamoto98a} comment that the difference in $z$ represents the only significant difference they observed in the features of dynamic heterogeneities between 2D and 3D. Where direct comparisons have been carried out~\shortcite{doliwa00,yamamoto98a}, there is no evidence that the long-wavelength anomalies that are a well documented feature of crystals and liquids in 2D have any significant impact on the glass transition in 2D. In this context, it is worth noting that the phenomenology of supercooled liquids in 3D and 4D is also very similar, with the most significant difference being that the breakdown of the scaling between the diffusion constant and the relaxation time (the Stokes-Einstein relation) is somewhat weaker in the higher dimension~\shortcite{eaves09}.

The kinetic length scales described in this Section - direct measures of the spatial distribution of particle displacements or bonds breaking - have proven valuable descriptive tools in establishing the reality of dynamic heterogeneities and their dependence on temperature. In Table \ref{summary}  we provide a summary of these approaches, identifying the time scales and/or lengths that must be chosen to resolve the transient heterogeneities in each case and the methods used to assign values to these quantities.  In an important development, explicit in box scaling and the displacement susceptibility approaches and implicit in the mobile cluster method, it becomes legitimate to select a quantity, not based on some particular physical justification, but rather on the pragmatic goal to maximise the resolution of the heterogeneity.

\begin{table}
\label{summary}

\begin{tabular}{|p{4.2cm}|p{3cm}|p{5cm}|}
\hline
\bf{Method} & \bf{Selected length/time scales} & \bf{Manner of Assigning Values} \\
\hline
\hline
box scaling$^{a}$ & threshold length & maximise kinetic length \\
\hline
mobile cluster$^{b}$ & observation time & maximise non-Gaussian parameter\\
& threshold length & maximum length $r^{*}$ for which $G(r^{*},t)=G_{gaus}(r^{*},t)$\\
\hline
bond breaking$^{c}$ & observation time & $0.5 \tau_{b}$ where $\tau_{b}$ is the average bond breaking time\\
& maximum~bond length & a length lying between the first two peaks of $g_{ij}(r)$\\
\hline
displacement-displacement correlation function$^{d}$ & observation time & maximise susceptibility $\kappa_{u}(t)$\\
\hline
\end{tabular}
\\
$^{a}$ \shortcite{hurley95}, $^{b}$ ~\shortcite{kob97}, $^{c}$ ~\shortcite{yamamoto98a}, $^{d}$ ~\shortcite{poole98b}\\
\\
\caption{ A summary of the various methods of calculating kinetic lengths and the associated length and time scales required in each case.}

\end{table}

Is there a \emph{best} measure of a kinetic length? There does appear to be a consensus concerning the choice of how to obtain a kinetic length. The kinetic length of choice is that associated with $\chi_{4}$, based on 4-point correlations and closely related to the displacement-displacement correlations introduced by Poole et al~\shortcite{poole98b}. As we shall see in the following Section, $\chi_{4}$ has a number of appealing features, not least being its close connection with the formalism of spin glasses. Comforting as such consensus can be, it is worth pointing out that $\chi_{4}$ provides the same information as that contained in the other measures reviewed in this Section. Its popularity depends, not on the superiority of its description of dynamic heterogeneities, but on its accessibility, particularly from experiments, and on its potential role in developing theoretical treatments. Whether $\chi_{4}$ provides a sufficiently complete account of cooperative dynamics is a question we shall return to in Section \ref{what}.

\section{Kinetic Lengths From 4-Point Correlations Functions}
\label{chi4}

A minimal description of dynamic heterogeneities requires that we measure the statistical correlations between the movement of pairs of particles. This description requires a correlation involving four positions: $\vec{r}_{1}(t)$, $\vec{r}_{1}(t+\tau)$, $\vec{r}_{2}(t)$ and $\vec{r}_{2}(t+\tau)$. The first calculation of such 4-point correlation functions from MD simulations of a Lennard-Jones mixture (similar to BMLJ2 except the size ratio $\sigma_{11}/\sigma_{22}= 5/8$) was reported by Dasgupta et al~\shortcite{dasgupta91} in 1991. The motivation of these calculations was to test for the presence of a growing length scale associated with fluctuations of the Edwards-Anderson order parameter, $\lim_{t \rightarrow \infty}<\delta n(\vec{r},t_{o})\delta n(\vec{r},t_{o}+t)>$,  where $n(\vec{r},t)$ is the density field. Such a growth in the length scale of an order parameter fluctuation would have provided evidence of a thermodynamic glass transition.  Fixing the spatial separation (i.e. $\vert \vec{r}_{1}-\vec{r}_{2}\vert$) at $2\sigma_{11}$, the authors found no evidence of correlated fluctuations at any time. It is likely that this negative result was a consequence of the choice of time interval. As we shall see, fluctuations of the density autocorrelation function exhibit a maximum at around $\tau_{\alpha}$, while the calculations of Dasgupta et al.~\shortcite{dasgupta91} were, by design, carried out in the plateau interval of the relaxation function, well short of this time scale.

In 1999, Franz, Donati, Parisi and Glotzer~\shortcite{franz99,donati02} described how the displacement-displacement correlations introduced to quantify the spatial distribution of particle movement (see previous Section) could be reformulated in terms of 4-point correlations \footnote{To get the chronology straight during this busy period, note that the paper \shortcite{donati02} first appeared as a preprint cond-mat/9905433 in 1999.} This reformulation involved a conceptual convergence of a method introduced to describe the spatial correlations of displacements with the formal tools developed to treat fluctuations in spin glasses, such as motivated Dasgupta et al~\shortcite{dasgupta91}. Beginning in 2000, Glotzer and coworkers~\shortcite{lacevic03,glotzer00b,lacevic02} presented a detailed account of the 4-point correlation function formalism as applied to supercooled liquids. Here we shall summarise their expressions~\shortcite{lacevic03} for a kinetic length scale using a 4-point correlation.

The structural relaxation is described by the quantity

\begin{equation}
\label{five}
Q(t)=\sum_{i=1}^{N}\sum_{j=1}^{N}w\vert(\vec{r}_{i}(0)-\vec{r}_{j}(t)\vert)
\end{equation}

\noindent a measure of the degree to which a configuration at time $t$ still overlaps the initial arrangement. Overlap of a particle with itself or another particle at an earlier time is established through the introduction of a window function $w(r)$ (where $w(r) = 1$ if  $\vert r \vert \le a$ and zero, otherwise). The self part $Q_{s}(t)$ of the relation function can be defined as

\begin{equation}
\label{six}
Q_{s}(t)=\sum_{i=1}^{N}w(\vert\vec{r}_{i}(0)-\vec{r}_{i}(t)\vert)
\end{equation}

\noindent La\v{c}evi\'{c} et al~\shortcite{lacevic03} showed that this self part is the dominant contributor to the relaxation function, the dynamic susceptibility and the kinetic length. Physically, this means that the essential relaxation event corresponds to the departure of a particle from its own initial `site'. Eq. \ref{six} provides an interesting link with the previous approaches~\shortcite{hurley95,kob97} in which the description of dynamic heterogeneities involved considering each particle moving beyond some threshold distance from its initial position. Similarly, Stein and Andersen~\shortcite{stein08,stein07} examine the 4 point correlations of a mobility defined as

\begin{equation}
\label{seven}
\mu(\vec{r},t)=\sum_{i=1}^{N_{A}}\delta(\vec{r}-\vec{r}_{i}(t))\mu_{i}(t)
\end{equation}

\noindent where $\mu_{i}(t)=1$ if $\vert\vec{r}_{i}(t+t^{*})-\vec{r}(t)\vert \ge d$ and zero otherwise.

Toninelli et al~\shortcite{toninelli05} and Flenner and Szamel~\shortcite{flenner07}, along with others, have used the (microscopic) self intermediate scattering function $F_{s}(q,t)$ instead of the self overlap function $Q_{s}(t)$. There is no fundamental difference between the two relaxation functions. The choice of the scattering wavevector $q$ in $F_{s}(q,t)$ selects the reference length scale analogous to the choice of the value of $a$ in the overlap function. The authors in ref. \shortcite{toninelli05} considered how various models of collective behaviour (elastic modes, domain wall fluctuations, etc.) where represented at the level of the 4 point correlations. Flenner and Szamel~\shortcite{flenner07} showed that the correlations of the fluctuations in the microscopic self intermediate function exhibited an anisotropy associated with the direction of particle motion relative to the vector between particle pairs.  This latter resulted expressed, in terms of the 4 point correlation function, the anisotropy that had previously been established by Doliwa and Heuer~\shortcite{doliwa00} who demonstrated that the kinetic length scale (obtained using Eq. \ref{two}) along the direction of the particle displacement was roughly twice as long (in a binary hard sphere mixture) as that along a direction perpendicular to the particle displacement.

A susceptibility $\chi_{4}(t)$, analogous to $\kappa_{u}(t)$ defined previously for the displacement correlations, is defined in terms of the fluctuations of the relaxation function $Q(t)$, i.e.

\begin{equation}
\label{eight}
\chi_{4}(t)=\frac{\beta V}{N^{2}}(<Q^{2}(t)>-<Q(t)>^{2})
\end{equation}

\begin{figure}[t]
\centering
\includegraphics*[width=8 cm]{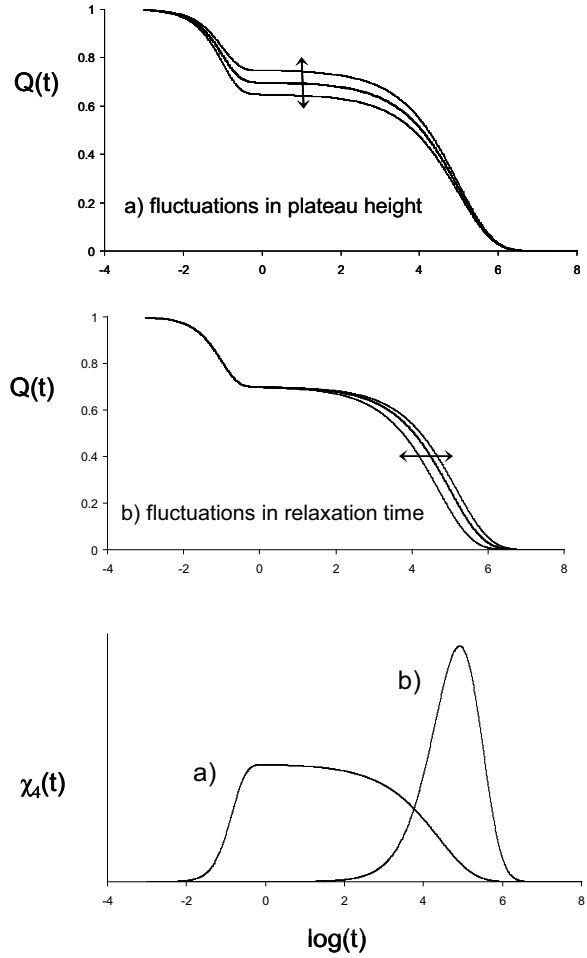}
\caption[]{Sketches of the fluctuations in $Q(t)$ due to a) fluctuations in the plateau height and b) fluctuations in the magnitude of the structural relaxation time. The time dependent dynamic susceptibilities $\chi_{4}(t)$, calculated using Eq.~\ref{six} for both types of fluctuations, are plotted in the bottom panel.
}
\label{figtwo}       
\end{figure}

\begin{figure}[t]
\centering
\includegraphics*[width=8 cm]{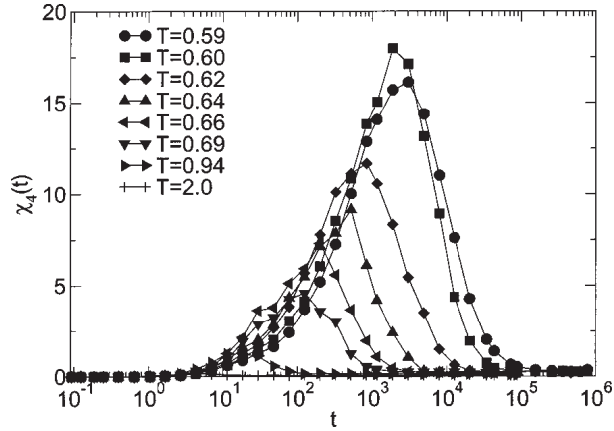}
\caption[]{The time and temperature dependence of $\chi_{4}(t)$ for the BMLJ2 mixture. The time corresponding to the maximum in $\chi_{4}(t)$ is found to be similar in value to the structural relaxation time $\tau_{\alpha}$ and to exhibit a similar T dependence. [Reprinted with permission from ref.~\shortcite{lacevic03} Copyright 2003, American Institute of Physics.]
}
\label{figthree}       
\end{figure}

\noindent The expression for the susceptibility $\chi_{4}(t)$ (Eq. \ref{eight}) in terms of the variance of the structural relaxation function $Q(t)$ is remarkable in that it provides information about the extent of dynamic heterogeneities (see below) without ever requiring that the dynamics be spatially resolved (as the methods in Section \ref{displacements} all did). Qualitative considerations of the fluctuations of $Q(t)$ also provide a useful way of differentiating the order parameter fluctuations, envisioned by Dasgupta et al~\shortcite{dasgupta91}, from the fluctuations in dynamics that are the subject of this review. In Fig.~\ref{figtwo} we present sketches of two distinct types of fluctuations in $Q$. The top panel represents fluctuations of the height of the plateau.  The Edwards-Anderson order parameter for spin glasses is this plateau height in an arrested system. An alternative type of fluctuation involves variations of the relaxation time (Fig.~\ref{figtwo}, middle panel). These two types of fluctuations are easily distinguished in their respective susceptibilities (Fig.~\ref{figtwo}, bottom panel). Where the plateau fluctuations result in a low amplitude $\chi_{4}(t)$, extended over the whole plateau time region, the dynamic fluctuations typically produce a more sharply peaked $\chi_{4}(t)$, with the maximum occurring at roughly the structural relaxation time. Compare these two (idealized) possibilities with the $\chi_{4}(t)$ calculated from simulations of the BMLJ2 mixture~\shortcite{lacevic03} in Fig.~\ref{figthree} and we find that fluctuations in dynamics, sketched as option b) in Fig.~\ref{figtwo}, provide a reasonable correspondence with the simulated liquid. The implication is that it is the dynamic fluctuations, as opposed to those of the plateau/order parameter, that dominate the observed susceptibility in the supercooled liquid. Kirkpatrick and Thirumalai~\shortcite{kirkpatrick88} were the first to point out that the fluctuations contributing to $\chi_{4}(t)$ could arise from these two distinct sources: order parameter fluctuations and fluctuations in the dynamics themselves.

\begin{figure}[]
\centering
\includegraphics*[width=8 cm]{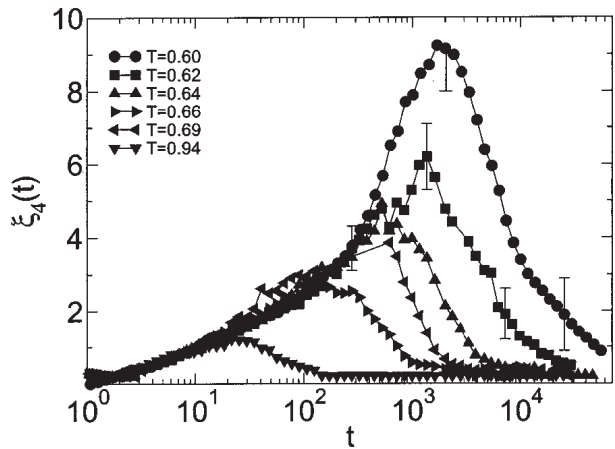}
\caption[]{The time and temperature dependence of the kinetic length $\xi_{4}(t)$ in the BMLJ2 mixture. [Reprinted with permission from ref.~\shortcite{lacevic03}. Copyright 2003, American Institute of Physics.]}
\label{figfour}       
\end{figure}

The susceptibility $\chi_{4}(t)$ can be directly obtained~\shortcite{lacevic03} from the 4-point correlation function via

\begin{equation}
\label{nine}
\chi_{4}(t)=\beta \int d\vec{r}g_{4}(\vec{r},t)
\end{equation}

\noindent where

\begin{equation}
\label{ten}
g_{4}(\vec{r},t)=\frac{1}{N\rho}\langle \sum_{ijkl}\delta(\vec{r}-\vec{r}_{k}(0)+\vec{r}_{i}(0))w(\vert\vec{r}_{i}(0)-\vec{r}_{j}(t)\vert)w(\vert\vec{r}_{k}(0)-\vec{r}_{l}(t)\vert)\rangle-\left<\frac{Q(t)}{N}\right>^{2}
\end{equation}

\noindent or

\begin{equation}
\label{eleven}
g_{4}(r,t)\equiv g_{4}^{ol}(r,t)-\left<\frac{Q(t)}{N}\right >^{2}
\end{equation}

\noindent where we have assumed that the correlations are isotropic and so only depend on the magnitude of the separation.  The significance of $g_{4}^{ol}(r,t)$ is that it corresponds to pair correlation function of those particles at $t = 0$ that end up overlapping with a particle at the later time $t$. It is this quantity, or rather its Fourier transform,

\begin{equation}
\label{twelve}
S_{4}^{ol}(\vec{q},t)=\int d\vec{r}g_{4}^{ol}(r,t)exp(-i\vec{q}\cdot \vec{r})\,
\end{equation}

\noindent that La\v{c}evi\'{c} et al~\shortcite{lacevic03} use to obtain the kinetic length $\xi_{4}$ whose dependence on $t$ is plotted in Fig.~\ref{figfour}. As emphasised in ref. \shortcite{lacevic03}, considerable care needs to be taken in ensuring that the 4-point correlator used to extract the kinetic length does not include a weak $O(1/N)$ tail associated with bulk fluctuations. As is generally the case, fluctuations, such as measured by the 4-point correlations, are very dependent on the choice of ensemble. Dalle-Ferrier et al~\shortcite{dalle-ferrier07} provide a useful discussion of this point.

Toninelli et al~\shortcite{toninelli05} have presented a thorough discussion of the behaviour of $\chi_{4}(t)$ over various time domains. The kinetic length increases in time as the heterogeneous character of the unrelaxed domains is exposed by the dynamics until it reaches a maximum, beyond which is decays to zero as the persistent domains finally succumb to relaxation and homogeneity is re-established. The temperature dependence of the maximum length $\xi_{4}(t^{*})$ in the BMLJ2 liquid is fitted by  $\xi_{4}(t^{*})=0.82(T/T_{c}-1)^{-0.82}$ with an apparent divergence at the mode coupling temperature $T_{c} ~ 0.57$~\shortcite{lacevic03}. Such apparent divergences generally do not actually eventuate and there is a considerable literature discussing the various interpretations of both the power law temperature dependence and the `avoidance' of the singularity~\shortcite{cavagna09}. As already mentioned, a scaling law, $\tau_{\alpha} \sim \xi_{4}(t^{*})^{z}$, was found to hold with $z = 2.34$.

The length scale $a$ imposed by the window function $w(r)$ plays a significant role in selecting the physical character of the fluctuations being measured. As pointed out in ref.~\shortcite{lacevic03}, select $a$ too small and the results are dominated by vibrational motions that obscure any heterogeneities, select $a$ too large and the notion of overlap quickly loses any physical significance as each particle can overlap with multiple particles. Dauchot et al~\shortcite{dauchot05} have chosen a value for $a$ the same way that $t^{*}$ is chosen, i.e. by finding the value that maximises $\chi_{4}$. Chandler et al~\shortcite{chandler06} pointed out that the use of a small $a$ (roughly, $a/\sigma \le 0.3$) results in the spatial clustering of immobile particles dominating the observed fluctuations, whereas the use of a large $a$ (i.e. $0.5 \le a/\sigma$) results in a $\chi_{4}(t)$ reflecting correlations in mobile particles. This length dependent crossover is related~\shortcite{chandler06} to the non-Fickian-to-Fickian crossover discussed in Section \ref{crossover}. Charbonneau and Reichman~\shortcite{charbonneau07} have described how the dependence of $\chi_{4}$ on the reference length scale differs when comparing liquids whose arrest is dominated by short-range repulsions and those dominated by short-ranged attractions.

The great attraction of the 4-point correlation function approach to the kinetic length is \emph{not} the 4-point correlation functions, $g_{4}(r,t)$ or $S_{4}(q,t)$, which are at least as difficult to use to calculate a kinetic length as any of the other methods described in the previous Section. The real appeal is in the quantity $\chi_{4}(t)$ and related susceptibilities. As the space integral of the 4-point correlation function (Eq.~\ref{seven}), $\chi_{4}(t^{*})$, the maximum susceptibility, represents a `volume of correlation' or, alternatively, a number $N_{corr}$ of correlated particles~\shortcite{dalle-ferrier07}. As the variance of the relaxation function $Q(t)$ (Eq.~\ref{eight}), $\chi_{4}(t)$ is no more difficult to calculate than the relaxation function itself. This latter virtue marks the superiority of $\chi_{4} (t)$ over the analogous susceptibility, $\kappa_{u}(t)$, from the displacement correlations (defined in Eq.~\ref{three}). Comparing Figs.~\ref{figthree} and \ref{figfour}, it is evident that $\chi_{4} (t)$ exhibits a qualitatively similar time dependence to that of the kinetic length $\xi_{4}(t)$ with a peak at some intermediate time, corresponding to a maximum in the differentiation of overlapping and non-overlapping domains.

This theoretical accessibility has been extended towards experimental accessibility in a series of papers~\shortcite{dalle-ferrier07,berthier05a,berthier07a,berthier07b,biroli06} starting with Berthier et al~\shortcite{berthier05a} in 2005. In ref.~\shortcite{berthier05a}, the authors showed how a lower bound on the value of $\chi_{4}(t^{*})$ could be obtained from 3-point correlations defined as the response of the structural relaxation function (i.e. $Q(t)$ or its analogue) to a change in a control parameter, such as temperature, pressure or density. This connection between a 3-point correlations and $N_{corr}$ has been explored in some detail~\shortcite{dalle-ferrier07}. Starting with the maximum of the 3-point susceptibility $\chi_{T}(t^{*})$ associated with the dynamic heterogeneities that are correlated to local enthalpy fluctuations, the number of correlated molecules is given (in the NPT ensemble) by

\begin{equation}
\label{thirteen}
N_{corr}= \sqrt{\frac{k_{B}T^{2}}{\Delta c_{P}}}max\{\vert\chi_{T}(t)\vert\}
\end{equation}

\noindent Eq.~\ref{thirteen} has been used to determine the size and temperature dependence of N$_{corr}$ for a range of molecular glass formers~\shortcite{dalle-ferrier07}. The discussion in ref.~\shortcite{dalle-ferrier07} of these results raises a number of important questions regarding kinetic lengths in general. The correlation volumes at $T_{g}$ were found to be modest in size. This result offers hope that simulations can contribute to the description of cooperative dynamics closer to $T_{g}$ than previously thought. It also cautions against arguments based on large separation of length scales in the supercooled liquid. Dalle-Ferrier et al~\shortcite{dalle-ferrier07} cast doubt on any simple connection between $N_{corr}$ and the cooperatively rearranging regions as imagined by Adam and Gibbs~\shortcite{adam65}. There is a quite basic difficulty in trying to translate observed correlations in mobility into the mechanism responsible for those correlations. We shall return to this point at the end of the review.

If both $\chi_{4}(t^{*})$ and $\xi_{4}(t^{*})$ can provide information about the extent of dynamic heterogeneities, what, exactly, is the connection between them? If one assumes that the spatial distribution of the mobile particles is scaled by a single length ($\xi_{4}(t^{*})$ in this case), it follows~\shortcite{biroli06,biroli04,whitelam04} that $\chi_{4}(t^{*})$ and $\xi_{4}(t^{*})$ are related by a power law,

\begin{equation}
\label{fourteen}
\chi_{4}(t^{*})=A(T)\xi_{4}(t^{*})^{2-\eta}
\end{equation}

\noindent Stein and Andersen~\shortcite{stein08,stein07} have confirmed the power law relation for the BMLJ1 mixture and found $\eta = -2.2$. More recently, Karmakar et al~\shortcite{karmakar10a} have carried out similar calculations but using a larger number of particles and reported a significantly smaller exponent, $\eta = -0.4$. Karmakar et al~\shortcite{karmakar10b} argue that the extraction of the kinetic correlation length from the 4-point structure factor $S_{4}(q,t)$ by fitting an Ornstein-Zernicke expression, i.e. $S_{4}(q,t)=\chi_{o}/(1+q^{2}\xi^{2})$, is inaccurate unless large systems (i.e. $N \sim 10^{5}$) are used.

Robust in definition and simple to apply, the susceptibility $\chi_{4}(t^{*})$ has proved a popular measure of the extent of cooperative motion. As obtained from the fluctuation of the 2-point correlation, $\chi_{4}(t^{*})$'s have been reported for colloids~\shortcite{duri06,cipelletti03,brambilla09,ballesta08}, granular material~\shortcite{dauchot05,keys07} and foams~\shortcite{mayer04}. In simulations, $\chi_{4}(t^{*})$ is being used to study cooperative behaviour in a wide range of systems - including those in non-Euclidean spaces and out of equilibrium.  Some examples of the latter applications: Sausset and Tarjus~\shortcite{sausset10} have calculated $\chi_{4}(t^{*})$  for a liquid of Lennard-Jones disks on a hyperbolic surface,  Furukawa et al~\shortcite{furukawa09a} have analysed the 4-point correlation functions in a liquid under steady shear (finding mobile regions to be elongated), and Parsaeian and Castillo~\shortcite{parsaeian08} have studied the effects of aging on $\chi_{4}(t^{*})$. Abraham and Bagchi~\shortcite{abraham08} have demonstrated that the low temperature magnitude of $\chi_{4}(t^{*})$ in a polydisperse Lennard-Jones mixture decreases with increasing width of the particle size distribution.

\section{Kinetic Lengths From Finite Size Analysis}
\label{finite}

\begin{figure}[t]
\centering
\includegraphics*[width=8 cm]{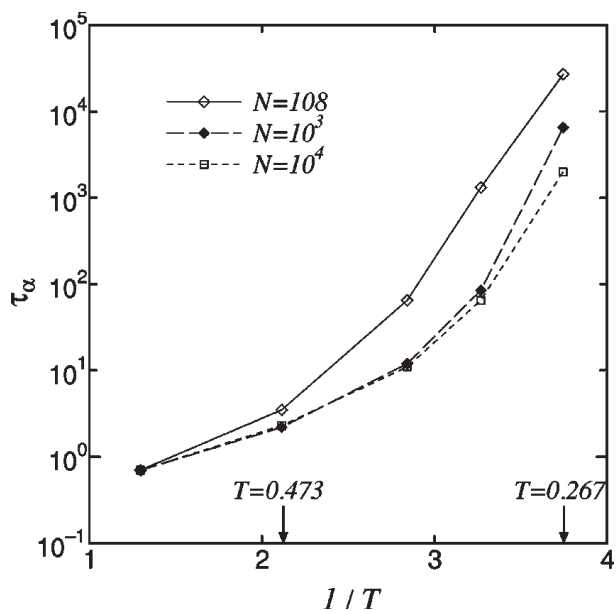}
\caption[]{The temperature dependence of $\tau_{\alpha}$ for $N = 108$ (open diamonds), $10^{3}$ (closed diamonds), and $10^{4}$ (open squares) for the SS mixture. [Reproduced with permission from ref.~\shortcite{kim00}.]
}
\label{figfive}       
\end{figure}

If heterogeneities are characterised by a length, then it follows that, as the size of a finite system approaches this inherent length, properties of the dynamics should show a dependence on the system size. For this finite size effect to be linked specifically to a kinetic length scale it is important to also establish that the static correlations in the liquid are independent of the system size over the size range studied. In 1992, Dasgupta and Ramaswamy~\shortcite{dasgupta92} reported the absence of any size dependence in the relaxation time of a density autocorrelation function for a binary Lennard-Jones mixture with a radius ratio of $5/8$. This study, however, failed to collect data for temperatures between $0.8T_{g}$ and $0.99T_{g}$, despite what appears to be an onset of a finite size dependence in this temperature interval. The authors, expecting to see the time scale \emph{decrease} with system size, interpreted the (slight) signs of the opposite trend as supporting their conclusion that there was no system size dependence. In 2000, Kim and Yamamoto~\shortcite{kim00}, studying the SS mixture for sizes $N = 108$, $1000$ and $10000$, found that as the temperature was lowered the relaxation time $\tau_{\alpha}$ exhibited a dependence on the system size, the relaxation time increasing as the system size decreased (see Fig.~\ref{figfive}). No dependence of the pair distribution function on system size was found for all temperatures studied.

The influence of system size on the relaxation times in the BMLJ1 mixture has been the subject of a number of studies. In 2003 Doliwa and Heuer~\shortcite{doliwa03} calculated the diffusion constants for four system sizes - $N = 65$, $130$, $260$ and $1000$ - down to $T_{c}$ ($\sim  0.43$). At the lowest temperature, they found $D_{65}/D_{130} \sim 1.2$ but with a large error ($ \pm 0.2$). They concluded that the size dependence was small, significantly less (and in the opposite direction) to that observed in the SS mixture~\shortcite{kim00}. Stein~\shortcite{stein07}, working with larger systems ($N = 1000$ and $8000$), found a slight decrease in the diffusion constant for the smaller system but, again, the difference was within the noise. In 2009, Karmakara et al~\shortcite{karmakar09} determined the relaxation time in a large number of systems across the size range $50 \le N \le 1600$ range of system sizes down to $T = 0.45$. They find a systematic increase in relaxation time with decreasing system sizes for $N$ (see Fig.~\ref{fignine}). The onset value of $N$ below which this finite size effect is observed increases from $\sim 100$ at $T = 0.8$ to $\sim 200$ at $T = 0.45$. These workers also studied the size dependence of  $\chi_{4}(t^{*})$ and, using a method first adapted from the study of critical phenomena to the glass problem by Berthier~\shortcite{berthier03}, determined a length scale from this data that increased on cooling from 2.1 (T=0.70) to 6.2 (T=0.45). An increase in the relaxation time with decreasing system size has also been reported in simulations of silica~\shortcite{zhang04,teboul06} but, unfortunately, there is no confirmation that the static properties remained unchanged, a nontrivial condition given the long range character of the potential in this model.

With the exception of ref.~\shortcite{doliwa03}, the data from simulations presents a picture of a modest but systematic increase in the volume associated with cooperative motion on cooling. Significantly, relaxation is \emph{slower} in small systems. We shall consider the implications and possible origin of this behaviour in Section \ref{what}. Confinement, whether imposed by walls or arising from inherent fluctuations in the supercooled liquid, appears to represent one generic mechanism for slowing down relaxation.

\section{Kinetic Lengths at Amorphous Interfaces}
\label{interface}

If your goal is to clarify a phenomenon in the homogeneous liquid, the inclusion of an interface is not usually a good idea. The problem is that the interface will typically perturb the liquid structure significantly and, thus, so alter the phenomenon from that found in the bulk as to obscure any connection with the homogeneous situation.  There is, of course, considerable interest in exactly such perturbed situations in the context of glass transitions actually taking place in confined geometries. This subject has been reviewed by a number of authors~\shortcite{binder05,baschnagel05}. Within the artificial world of simulations, however, it is possible to contrive a surface that is structurally neutral by simply freezing the motions of some portion of the liquid. Such walls are an example of the imposition of a kinetic constraint. The interesting question is then to determine the length scale over which the influence of such a constraint is propagated into the unconstrained liquid. The idea of using an interface to establish the kinetic length scale was described~\shortcite{butler91} in 1991 in the context of a lattice model of glassy kinetics.

Scheidler et al~\shortcite{scheidler00,scheidler04} have reported simulation studies of the BMLJ1 mixture adjacent to a rough wall made up of the frozen liquid. These workers have considered both a cylindrical pore~\shortcite{scheidler00} and a planar wall~\shortcite{scheidler04}. While the idea of a frozen liquid wall is simply sketched, its implementation takes some care.  In ref.~\shortcite{scheidler04} the temperature of the liquid used to produce the frozen walls was adjusted to minimize any structural perturbation and an additional repulsive potential was included to prevent particle penetration into the wall. Scheidler et al~\shortcite{scheidler04} fitted the relaxation time $\tau_{q}(z)$ (associated with the decay of a self intermediate relaxation time $F_{s}(q,z,t)$ at a distance $z$ from the wall) for the BMLJ1 mixture to three different functions of the normal distance $z$ from the planar wall. The best fit was obtained with the following expression,

\begin{equation}
\label{fifteen}
\ln\left[\frac{\tau_{q}(z)}{\tau_{o}}\right]=A(T)\exp\left[-\frac{z}{\xi_{o}(T)}\right]
\end{equation}

The kinetic length $\xi_{o}$ increased by a factor of 3 as the temperature dropped to $T = 0.5$.  The increase in $\xi_{o}$ with temperature was fitted with a simple Arrhenius form. The temperature dependence of $\xi_{o}$ appears to be quite different from the $(T-T_{c})^{-\gamma}$ dependence reported for the length of the dynamic heterogeneities in the bulk for the same model~\shortcite{donati99a}.  Obtaining values of the surface kinetic length at temperatures closer to $T_{c}$ have been frustrated by poor statistics.

Where Scheidler et al~\shortcite{scheidler00,scheidler04} have described how effective a frozen liquid is in imparting its immobility to an adjacent mobile liquid, Cavagna et al~\shortcite{cavagna07,biroli08} have examined how the frozen liquid can actually constrain the configurations available to the mobile liquid. The use of amorphous boundaries to establish an equilibrium length scale associated with structural correlations was described by Bouchaud and Biroli~\shortcite{bouchaud04} and Montanari and Semerjian~\shortcite{montanari06}. Instead of a frozen half plane, Cavagna et al~\shortcite{cavagna07} have considered a spherical shell of frozen SS liquid, enclosing the mobile liquid in a volume of radius R. We emphasise that this study does not deal with the kinetics of the confined liquid but, instead, uses an amorphous wall to extract a \emph{static} length scale. It is included in this review of kinetic lengths, in part, because of the obvious methodological parallels with the work of Scheidler at al~\shortcite{scheidler04}. There is also a general expectation that kinetic length scales derive from underlying static length scales. Mark Ediger~\shortcite{ediger00} expressed the situation with admirable delicacy, ``\emph{At present, it is an article of faith that something in the structure is responsible for dynamics that can vary by orders of magnitude from one region of the sample to another at $T_{g}$}''. Recent work~\shortcite{cammarota10} has suggested that the origin of the growing kinetic length scale might lie in a separation of phases characterised by distinct degrees of the amorphous order. In ref.~\shortcite{biroli08}, the authors determine, using an accelerated Monte Carlo algorithm, the degree of overlap $q_{c}$ of the ensemble of configuration at the centre of the sphere as a function of the radius R. They found that the overlap decayed to the random value $q_{o}$ as

\begin{equation}
\label{sixteen}
q_{c}(R)-q_{o}=\Omega \exp[-(r/\xi)^{\zeta}]
\end{equation}

\noindent with both the length $\xi$ and the exponent $\zeta$ increasing with supercooling - at $T = 0.482$ the length $\xi =0.617$ and the exponent $\zeta \sim 1$ and both increase on cooling so that at  $T = 0.203$, $\xi =3.82$ and $\zeta = 4.00$.

An earlier study of a liquid confined within a frozen liquid shell was presented by Sim et al~\shortcite{sim98,sim99}. These authors studied a single component Lennard-Jones liquid in 2D where the crystallization was frustrated by the disorder of the frozen wall. They did not present any systematic results associated with a kinetic length. However, the model is of potential interest due to the simplicity of the confined liquid and the possibility of an explicit counting of allowed configurations. The model represents an extension of the classic problem of packing disks in finite containers~\shortcite{desmond09}. Sim et al~\shortcite{sim99} reported some specific examples of collective reorganization events.

The geometry of the pinned particles - planes, tubes or spherical cavities - reflect the particular problem that inspired the authors. In the case of Scheidler et al~\shortcite{scheidler04}, this was the influence of confinement on glassy dynamics, while for Cavagna et al~\shortcite{cavagna07} it was to test the idea of droplet excitations as conceived in the mosaic theory~\shortcite{kirkpatrick87,kirkpatrick89,xia01,lubchenko03,bouchaud04}.  What about a spatial distribution of pinned particles that does not impose any particular correlation? Supercooled liquids subject to random pinnings have been studied by a number of groups~\shortcite{kim03,lin06}. In 2003 Kim~\shortcite{kim03} reported the effect of pinning the positions of a fixed number $N_{d}$ of randomly selected particles in the SS mixture. He found that the relaxation time $\tau_{\alpha}$ scaled with $N_{d}$ and $T$ as

\begin{equation}
\label{seventeen}
\tau_{\alpha}(T,N_{d})\propto \exp[N_{d}/T^{\nu}]
\end{equation}

\noindent with $\nu = 3.7$. In trying to extract a kinetic length scale from this calculation, Kim resorted to the following argument. There is one length scale imposed by the defects, i.e.

\begin{equation}
\xi_{d}(T) \propto (V/N_{d})^{1/3}\,
\end{equation}

\noindent and there is another, the intrinsic kinetic length $\xi (T)$, the one we are actually interested in. If one assumes that a) the intrinsic kinetic length is unperturbed by the pinned particles and b) that at the glass transition (defined as $\tau_{\alpha}$ equalling some big number) $\xi_{d}(T)=\xi (T)$, then it follows that

\begin{equation}
\label{eighteen}
\xi (T) \propto T^{-\nu/3}
\end{equation}

\noindent The argument is awkward. In particular, assumption b) above neglects the crossover to simple unpinned behaviour when $\xi_{d} (T) > \xi (T)$. The methodology, however, has potential as a general tool for exploring kinetic and static correlations through the imposition of dilute random pinnings.

\section{Kinetic Lengths from Cross-Over Behaviour}
\label{crossover}

We shall complete our survey of length scales with those that arise most directly from the dynamical processes of interest and, therefore, perhaps represent the most pressing demand for our attention. The length scales described in this Section take as their starting point the existence of some sort of length dependent crossover in a physical property explicitly associated with particle motion.  The existence of dynamic heterogeneities leads, not surprisingly, to a range of physical behaviour that deviates from that expected of a uniform system. As one probes the behaviour over length scales larger than that of the heterogeneity, the `classical' behaviour is recovered and, accordingly, a length scale can be associated with this crossover to homogeneity.

Self (tracer) diffusion is one such phenomenon that exhibits a crossover that is manifest in changes in the time dependent displacement probability distribution $G(r,t)$ (the van Hove distribution function). The story around $G(r,t)$ in glass forming liquids is quite rich and so we shall take a moment to sketch out some main points, at least to clarify some terminology. In 1990, Odagaki and Hiwatari~\shortcite{odagaki90} noted that, at fixed times, $G(r,t)$ underwent a transition from Gaussian to non-Gaussian as a liquid was supercooled. Hurley and Harrowell~\shortcite{hurley96} pointed out that this was an expected consequence of the increase in dynamic heterogeneity; specifically, the presence of persistent kinetic subpopulations. Dynamic heterogeneities have also been invoked to resolve another puzzle involving self diffusion. Sillescu and coworkers~\shortcite{fujara92} had shown experimentally that, on supercooling, fragile liquids exhibited a breakdown in the scaling between the diffusion constant $D$ and both the shear viscosity (the Stokes-Einstein relation) and the rotational diffusion constant (the Debye expression). This phenomenon is actually evident just among the different length scales of the self intermediate scattering function $F_{s}(q,t)$,  the Fourier transform of $G(r,t)$. The temperature dependence of the relaxation time of $F_{s}(q,t)$ exhibits, on supercooling, an increasing anomalous dependence on $q$, with the small $q$ time scale behaving like $D$ (as it must) and the large $q$ relaxation exhibiting a temperature dependence similar to that of the shear viscosity~\shortcite{perera98a,chandler06,chaudhuri07}. This general loss of a single time scale on supercooling is referred to as `decoupling'. A number of groups~\shortcite{cicerone96,berthier04,berthier05b} have reached the following consensus regarding the origin of this decoupling. The idea is that different transport properties correspond to different moments of the distribution of microscopic times, so their decoupling at a particular wave vector is associated with the growth of dynamical heterogeneity (as manifest in the broadening of the distribution of microscopic times) over the corresponding length scale. Chaudhuri et al~\shortcite{chaudhuri07} have pointed out that the presence of an exponential tail in the van Hove function is a signature of the presence of slow and fast particles and can account for the decoupling of diffusion and structural relaxation. The continuous time random walk model they propose has been further quantified by Hedges et al~\shortcite{hedges07} who demonstrated that the ratio of the persistence time over the exchange time (the times for the first move and between subsequent moves, respectively) grows rapidly in the supercooled liquid.

The explanation of decoupling provided in refs.~\shortcite{chaudhuri07,hedges07} requires that the different transport processes can be related to a common distribution of microscopic times. While this condition is met for the relaxation of $F_{s}(q,t)$ at different $q$, it is not clear that it is met by the Stokes-Einstein breakdown itself since diffusion and viscosity correspond to quite different physical processes and, therefore, are associated with quite distinct distributions of microscopic times. Other approaches~\shortcite{hodgdon93} to the decoupling of diffusion and viscosity avoid this particular criticism by retaining explicit coupling between local mobility fluctuations and stress relaxation.

\begin{figure}[t]
\centering
\includegraphics*[width=8 cm]{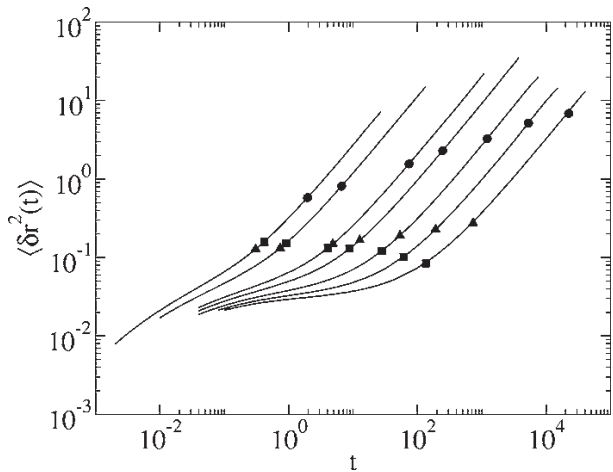}
\caption[]{The time dependence of the mean square displacement for the A particles in the BMLJ1 model for T = 1.0, 0.8, 0.6, 0.55, 0.50, 0.47 and 0.45 listed from left to right. The symbols are placed at different characteristic times. Squares: the time at which the standard non-Gaussian parameter reaches the maximum value. Triangles: the $\alpha$ relaxation time $\tau_{\alpha}$. Circles: the onset time for Fickian diffusion. [Reproduced with permission from ref.~\shortcite{szamel06}.]
}
\label{figsix}       
\end{figure}

Over a long enough time, particles will sample a sufficient number of kinetic environments so as to recover standard or Fickian diffusion. Szamel and Flenner~\shortcite{szamel06} have determined the time $\tau_{F}$ over which a particle must, on average, move before $G(r,t)$ becomes Gaussian for the BMLJ1 mixture simulated using Brownian dynamics. Their results are shown in Fig.~\ref{figsix}. They observe that $\tau_{F}$ is larger than the relaxation time $\tau_{\alpha}$ and that this difference increases as the temperature is lowered. A length scale can be obtained from Fig.~\ref{figsix} simply by reading off the mean square displacement at $t = \tau_{F}$. This length $l^{*}$ ranges up to $~ 2.5$ small particle diameters and corresponds to the distance that a particle must on average move before exhibiting Fickian diffusion. The value of $l^{*}$ from ref.~\shortcite{szamel06} is roughly half that predicted by the expression for the crossover length due to Berthier et al~\shortcite{berthier05c}, $l^{*} \propto \sqrt{D\tau_{\alpha}}$, but, at large supercoolings, the two lengths exhibit a similar temperature dependence.

Stariolo and Fabricius~\shortcite{stariolo06}, in a study of the BMLJ1 mixture, reported the appearance of a new crossover in the self intermediate scattering function at around $\tau_{\alpha}$, in addition to the crossover to Fickian diffusion at longer times. The authors associated the earlier crossover with the length scale of the dynamic heterogeneities as measured by $\chi_{4}$. A link between the length scale of the crossover to Fickian behaviour and the length scale of the dynamic heterogeneities as obtained from the 4 point correlations has been explored by Berthier~\shortcite{berthier04}. Studying the BMLJ1 model, he demonstrated that a suitably normalised product of the q-dependent relaxation time and the diffusion constant from a wide range of temperatures could be collapsed onto a single master curve when the wavevector $q$ was scaled by the length scale of dynamic heterogeneities.

Is the crossover to Fickian behaviour governed by a time scale $\tau_{F}$ or a length scale $l^{*}$? Given the transient character of dynamic heterogeneities, the answer appears to be a matter of taste. A crossover involving an unequivocally \emph{static} distribution of heterogeneities has been studied by Barrat and coworkers~\shortcite{leonforte05}. These authors demonstrated that the continuum elastic description of a disordered polydisperse mixture of Lennard-Jones particles at zero temperature broke down for length scales less than $\sim 23$ particle diameters. Over length scales smaller than this crossover length, the material exhibited non-affine response to an applied strain and this length gives the lower wavelength bound for the applicability of classical eigenvectors.

The connection between non-affine displacements and the approach to the glass transition has been made explicit in a study by Mosayebi et al~\shortcite{mosayebi10}. The local potential energy minima of the BMLJ1 model have been collected from MD trajectories as a range of temperatures. These inherent structures are subjected to a static shear deformation at $T = 0$ and the spatial distribution of the resulting non-affine displacements calculated. These authors find that a characteristic length scale of the non-affine field grows as the temperature from which the inherent structures are obtained is lowered.

While the complexities of normal modes of disordered materials has not traditionally been associated with dynamic heterogeneities, a growing body of evidence suggests that the two phenomena are correlated~\shortcite{schober93,oligschleger99,brito06,brito07,widmer-cooper08,widmer-cooper09}. It is possible that the large, but finite, length scales identified in ref.~\shortcite{leonforte05} represent a useful $T = 0$ limit for the length scale of dynamic heterogeneities.

\begin{figure}[t]
\centering
\includegraphics*[width=8 cm]{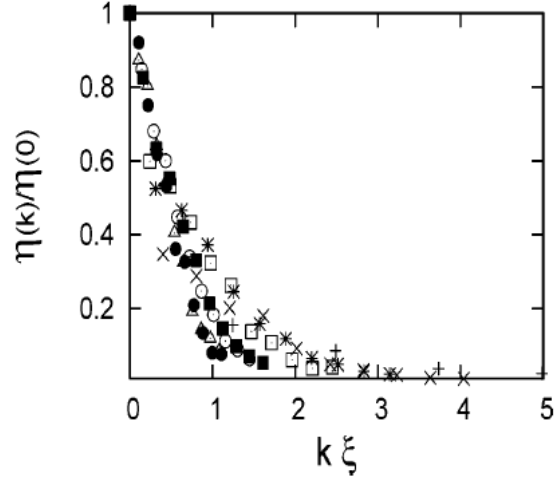}
\caption[]{The wavevector-dependent viscosity for the BMLJ1 mixture divided by the $k = 0$ value, vs the reduced wavevector for 200 K, 160 K, 120 K, 100 K, 80 K and 60 K. The value of $\xi$, the kinetic length, is obtained from the fitting function in Eq.~\ref{nineteen}.  [Reproduced with permission from ref.~\shortcite{kim05}.]}
\label{figseven}       
\end{figure}

So far in this review we have made no mention of a length scale associated specifically with stress relaxation. Certainly, the overwhelming emphasis of simulation studies of the glass transition has been on single particle dynamics, in spite of the central role of viscosity in defining the glass transition. There are, however, a number of interesting papers on growing length scales associated with transverse momentum fluctuations, the kernel of the Green-Kubo expression for shear viscosity. In 1995, Mountain~\shortcite{mountain95} used the transverse momentum autocorrelation function to obtain the dispersion curve for the supercooled SS mixture. He identified a growing length scale as the longest wavelength associated with propagating transverse modes. This wavelength, obtained by extrapolating the dispersion curve for the supercooled liquid to the point where the mode frequency vanished, showed a strong increase at large supercoolings. At the temperature at which Yamamoto and Onuki~\shortcite{yamamoto98a} found a kinetic length of $\sim 10$, Mountain reports a length of $\sim 55$ ( a value roughly three times the length of his own simulation box). Ahluwalia and Das~\shortcite{ahluwalia98}, using a mode coupling theory, have argued that the length identified by Mountain will diverge as $T_{g}$ is approached from above.  There is a serious difficulty in associating the length described by Mountain and Das with the kinetic length scales that are the subject of this review.  As pointed out by Hiwatari and Miyagawa~\shortcite{hiwatari90}, a straightforward application of viscoelastic theory results in the prediction that the longest wavelength associated with transverse propagation is proportional to $\eta$, the shear viscosity.  This relation simply reflects the condition that, for a mode to propagate, its frequency must exceed $1/ \eta$ , the value set by the dissipation. The growing length described by Mountain~\shortcite{mountain95} is a direct consequence of the growing relaxation time and is quite independent of any microscopic correlation length associated with the physical origin of this growing relaxation time.

An alternate and more informative treatment of the transverse momentum correlation function has been presented by Kim and Keyes~\shortcite{kim05}. These workers have calculated the time integral of the k-dependent transverse momentum autocorrelation function. This quantity, essentially the zero frequency component of the correlation function, is the wavevector-dependent shear viscosity $\eta (k)$. The authors~\shortcite{kim05} found that $\eta (k)$, calculated for the BMLJ1 model, could be fitted with the following functional form,

\begin{equation}
\label{nineteen}
\frac{\eta (k)}{\eta (0)}=1+atanh(k\xi (T))
\end{equation}

Consistent with Eq.~\ref{nineteen}, $\eta (k)/ \eta (0)$  plotted against $k \xi (T)$ collapses the data from different temperatures onto a single master curve (as shown in Fig.~\ref{figseven}) that decays with increasing wavevector. The length scale $\xi$ was found increase from 0.13 to 1.62 $\sigma_{AA}$ on cooling. Kim and Keyes~\shortcite{kim05} go on to argue that the increase in $\xi$ can be directly linked to the breakdown in the Stokes-Einstein scaling through the use of a mode coupling expression due to Keyes and Oppenheim~\shortcite{keyes73}.

Furukawa and Tanaka~\shortcite{furukawa09b}, studying the SS model, have extended the Kim-Keyes observations in two ways. First, they have shown that the increasing wavevector-dependence of $\eta (k)$ is due entirely to the transverse component of the momentum flux and, second, they have established (again, for the SS mixture) that the length extracted from $\eta (k)$ has the same magnitude and temperature dependence as $\chi_{4}$ obtained, as described in Section \ref{chi4}, from fluctuations in the structural relaxation function. Puscasu et al~\shortcite{puscasu10a,puscasu10b} have simulated the wavevector dependence of the velocity kernel for a range of liquids including diatomic molecules and short chain alkanes and, for the latter model, report the growth of a very large length scale as the supercooling is increased.

The decay of $\eta (k)$ with wavevector $k$ reflects the decrease in dissipation as the wavelength shortens. As harmonic solids show no dissipation, it is tempting to associate the growing length identified in refs.~\shortcite{kim05,furukawa09b} as being associated with the characteristic length scale of such solid-like domains. The (rough) superposition of the normalised $\eta (k)$'s from both high and low temperatures, as shown in Fig.~\ref{figseven}, suggests that, in the supercooled liquid, stress relaxes similarly to that in the high temperature liquids except that the elementary objects are now rigid clusters with a linear dimension $\xi$ instead of the atomic components. In 1989, Ladd and Alder~\shortcite{ladd89} described the stretched tail of the shear stress autocorrelation function in hard sphere liquids near freezing. (Their evocative label - the `molasses' tail - does not seem to have caught on.) Subsequently, Isobe and Alder~\shortcite{isobe09,isobe10} have argued that the long time relaxation of the shear stress is dominated by the life time of rigid clusters in the liquid.  

\section{What Lengths Influence Relaxation?}
\label{what}

\begin{figure}[t]
\centering
\includegraphics*[width=8 cm]{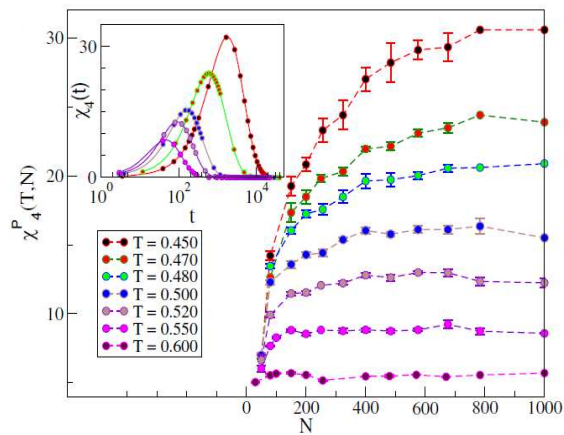}
\caption[]{Peak height of the dynamic susceptibility, $\chi_{4}^{P}(T,n)$ for the BMLJ1 model plotted as a function of system size $N$ for different temperatures. For each temperature, $\chi_{4}^{P}(T,n)$ increases with system size, and saturates for large system sizes. $\chi_{4}^{P}(T,n)$ also increases as the temperature is lowered. Insert: $\chi_{4}^{P}(T,n)$ plotted as a function of time. In the main plot, temperature increases from the top curve to the bottom. In the insert, temperature decreases moving from the left curve to the right.[Reproduced with permission from ref.~\shortcite{karmakar09}.]
}
\label{figeight}       
\end{figure}

\begin{figure}[t]
\centering
\includegraphics*[width=8 cm]{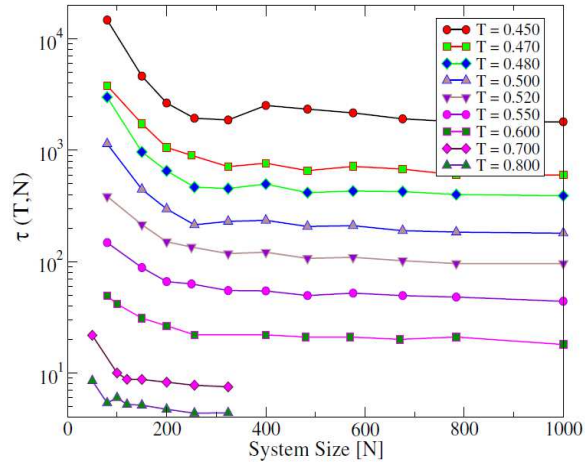}
\caption[]{Relaxation times as a function of temperature and system size in the BMLJ1 mixture. For the smallest temperature, $\tau (T,N)$ increases by approximately a decade from the largest to the smallest system size. Temperature increases from the top curve to the bottom.[Reproduced with permission from ref.~\shortcite{karmakar09}.]
}
\label{fignine}       
\end{figure}

Does a kinetic length provide the unified description of cooperative dynamics for which it was intended?  We have seen that the relaxation time of a glass forming system can be increased by either decreasing the temperature or decreasing the system size below some threshold value. Karmakar et al~\shortcite{karmakar09} have examined the dependence of both the relaxation time $\tau$ and the susceptibility $\chi_{4} (t^{*})$  ($\chi_{4}^{P}$ in their notation) of the BMLJ1 mixture as a function of temperature and number of particles. Their results are shown in Figs.~\ref{figeight} and \ref{fignine}.  Fixing the system size at a large value, say $N=1000$, we see that $\chi_{4}^{P}$ and $\tau$ both increase as T decreases, similar to behaviour already described in Section \ref{chi4}. If, however, we hold T fixed, the variation of $\chi_{4}^{P}$ and $\tau$  with respect to $N$ have opposite signs. The authors note that this result is contrary to the expectations of finite scale scaling. It indicates that $\chi_{4}^{P}$  does not contain all of the information about the collective processes in the liquid necessary to establish the relaxation time.

The puzzling observations of Karmakar et al~\shortcite{karmakar09} had been foreshadowed by earlier work.  Kim and Yamamoto~\shortcite{kim00} demonstrated that the increase in the relaxation time they observed for the small system was accompanied, not unexpectedly, by a truncation of the size of mobile clusters. The point here is that the more extended mobile regions may represent greater mobility for the system, not less. J\"{a}ckle and coworkers~\shortcite{frobose00,jackle01} demonstrated that, even in systems exhibiting dynamic heterogeneities, the diffusion of particles still involved significant coupling to extended visco-elastic flows (i.e. collective motions in which particles retain their neighbours) dominated by the longest wavelength transverse modes in the system. Decreasing the system size removes the longer wavelength modes and, possibly, stiffens the surrounding medium, leading to a decrease in mobility. Such a scenario would be expected to exhibit a very different system size dependence to the way size influenced the dynamic heterogeneities themselves. The spatial correlations in small displacements, as opposed to the large ones that typically define `mobile' particles, are an important component of the length that characterises relaxation, both incoherent (e.g. self diffusion) and coherent (e.g. stress relaxation). In a recent study of structural relaxation~\shortcite{widmer-cooper09}, it was shown that, of the particle movements that have contributed irreversibly to relaxation, 60\% (at the lowest temperature studied) could be categorized as strain-like, meaning that they involved the loss of no more than one of the initial neighbours.

\section{Conclusions}
\label{section:concl}

The measures of a kinetic length scale reviewed in this chapter have succeeded in a number of things. They have confirmed, by the act of measurement, the existence of spatial heterogeneity of the kinetics and the coarsening of this distribution on cooling. This descriptive success is an important milestone. It allows for the comparison between different glass formers, between the spatial character of the dynamics and that of the static properties of the supercooled liquid and between various theoretical treatments of the glass transition. The repeated observation of power law relations between a kinetic length and a relaxation time is significant. Such behaviour is suggestive of relaxation processes governed by the fluctuations in domain size. An alternative, in which the length corresponded to the dimension of an object involved in an activated process, would be expected to exhibit a different relation~\shortcite{cavagna09}, i.e.

\begin{equation}
\label{twenty}
\tau \propto \exp\left[\frac{A\xi^{\zeta}}{k_{B}T}\right]
\end{equation}

The power law exponents have been found to lie, roughly, between 2 and 4. If we think of the glass transition as being the temperature at which a relaxation time as increased by a factor of, let us say, $10^{10}$, then these power laws would require the kinetic length at this glass transition to have increased by a factor of between $10^{2}$ and $10^{5}$. The reports of temperature dependences for the kinetic length with singularities at the mode coupling temperature are generally regarded as a signature, not of a divergence, but of a cross-over to some alternative relaxation mechanism. To date there is no evidence for the large increases in kinetic lengths suggested by the observed power law. The conclusion is then that the time-length relationship extracted over a limited range of temperatures in simulations does not continue to hold at lower temperatures. The slow down in the growth of the kinetic length on cooling has been noted in experiments~\shortcite{dalle-ferrier07,brambilla09}. In simulations of the BMLJ1 mixture, Berthier et al~\shortcite{berthier07b} reported that for $T \le 0.47$, the growth of the dynamic susceptibility with respect to $\tau_{\alpha}$ becomes much slower than that observed at higher temperatures, ``perhaps logarithmically slow''. This last comment from \shortcite{berthier07b} is a reference to the prediction of a logarithmic relation $\xi \sim (\ln \tau_{\alpha})^{\zeta}$ , such as expressed in Eq.~\ref{twenty}, from theories that invoke an activated process~\shortcite{lubchenko03,bouchaud04}.

The kinetic lengths reviewed in Sections \ref{displacements} and \ref{chi4} all provide useful and, essentially similar, measures of the dynamic heterogeneities. An important question is, however, how well can they account for the length scales implicit in finite size effects, interfacial correlations and crossover behaviour? Furukawa and Tanaka~\shortcite{furukawa09b} have reported that the characteristic length obtained from the q-dependence of the transverse momentum fluctuations exhibits a similar size and temperature dependence to $\chi_{4} (t^{*})$. Berthier's~\shortcite{berthier04} demonstration that the Fickian crossover can be scaled by the length scale from $S_{4}(q,t)$ establishes a similar link. On the other hand, the increase in time scale due to the reduction in system size points to the role of correlations not included in the dynamic susceptibility. There remains an open problem to establish a clear identification of an explicit kinetic length (i.e. analogous to the lengths defined in Sections \ref{displacements} and \ref{chi4}) with an implicit length scale such as demonstrated in Sections \ref{finite}, \ref{interface} and \ref{crossover}.

As already touched upon, there is a gap between description and mechanism, one that only becomes truly evident now that the problem of description has been largely solved. To date, it is not clear that dynamic heterogeneities have clarified the mechanisms of relaxation. Part of the challenge in addressing the issue of mechanism certainly lies in refining what it is we actually want explained. The growth of the activation energy on cooling fragile liquids is generally associated with the growth of the number of elementary processes that must occur in series to achieve relaxation.  There is certainly clear evidence, through the finite size results, amorphous interface studies and the observation of the characteristic length in the transverse momentum fluctuations, that the mechanisms responsible for relaxation of particle positions and stress do have characteristic lengths and that these grow as the temperature decreases. Missing is a description of those elementary process by which the observed length scales are generated.

There remain a number of interesting and open challenges that arise directly from what we have already learnt from MD simulations. We shall conclude by listing a few. First, there is evidence that the number of mobile particles (however one might define them) in a supercooled liquid undergoes substantial fluctuations in time.  The democratic particle approach, introduced by Appignanesi and coworkers~\shortcite{appignanesi06a,appignanesi06b}, provides one explicit measure of the intermittent fluctuations in the number of particles involved in large displacements. From experiments on granular systems~\shortcite{candelier09} and, subsequently, MD simulations~\shortcite{candelier10}, Dauchot and coworkers have identified the intermittent appearance of spatially correlated bursts of enhanced mobility among particles. Christened `avalanches', this intermittent behaviour appears to increase as the glass transition is approached.  Clarification of the significance of these fluctuations is important. Is there a new length scale (or a hierarchy of new lengths) describing the volume over which the avalanche occurs? Is relaxation at low temperatures increasingly dominated by intermittent bursts of activity and, if so, what is happening during the quiescent intervals that leads to initiating a mobility event?

A second challenge is to make direct connection between dynamic heterogeneities and coherent processes like shear stress relaxation. The growing length scale associated with the transverse momentum fluctuations~\shortcite{kim05,furukawa09b,puscasu10b} provides a starting point. Along with particle mobility, dissipation is also becoming increasingly heterogeneously distributed as the temperature drops. To understand such phenomena we need to understand how to construct a description of visco-elastic behaviour where the dissipation can be localised even as the elastic behaviour becomes more extensive with the approach to the glass state.

Finally, it remains a fundamental tenant that the ultimate origin of a kinetic length scale lies in length scales associated with structural correlations. Understanding this link between structure and dynamics is a problem that, currently, only simulations can address. There are no shortage of aspirants for the missing link: the spatial distribution of localised soft modes~\shortcite{schober93,oligschleger99,brito06,brito07,widmer-cooper08,widmer-cooper09}, the mosaic length scale~\shortcite{cavagna07,biroli08}, the physical extent of clusters of locally preferred structures~\shortcite{coslovich09,tanaka10,lerner09,mossa06,pedersen10} to name some. Should the relevant structural length scales ever be successfully unearthed it is quite possible that the description of dynamic heterogeneities may be rendered irrelevant, replaced by the more convenient and illuminating account provided by the relevant structure. That dynamic heterogeneitiy may prove the agent of its own demise (as a descriptor of cooperativity, that is) is a real possibility. Until then, the spatial distribution of dynamics remains our most general and concrete description of the complex dynamics associated with a liquid's passage to rigidity in the glass state.

\section{Acknowledgements}

I gratefully acknowledge valuable conversations with David Reichman, Grzegorz Szamel, Gilles Tarjus and Peter Daivis. I would also like to thank Luca
Cipelletti for his generous help in preparing the manuscript. This work has been supported in part by the Australian Research Council through the Discovery
Grant program.

\bibliographystyle{OUPnamed_notitle}
\bibliography{refs_chap7_corrected}
\end{document}